\documentclass[useAMS,usenatbib,usegraphicx]{mn2e}
\usepackage[english]{babel}
\usepackage{times}
\usepackage[applemac]{inputenc}
\usepackage{amsmath}
\usepackage{fancybox}
\usepackage{fancyhdr}
\usepackage{color}
\usepackage{amssymb} 
\usepackage{inputenc} 
\usepackage{enumerate}
\usepackage{graphicx}

\newcommand{\apj}{ApJ}

\newcommand{\apjl}{ApJL}
\newcommand{\mnras}{MNRAS}
\newcommand{\aap}{A\&A}

\def\4u{4U~0614+09}
\begin{document}

\title{On the distribution of frequency ratios of kHz QPOs}
\author[Martin Boutelier, Didier Barret, Yongfeng Lin \& Gabriel T\"or\"ok]{Martin Boutelier$^{1,2}$\thanks{Email : martin.boutelier@cesr.fr}, Didier Barret$^{1,2}$ \& Yongfeng Lin$^{1,2,3}$ \& Gabriel T\"or\"ok$^4$\\
$^1$Université de Toulouse (UPS) \\
$^2$Centre National de la Recherche Scientifique, Centre d'Etude Spatiale des Rayonnements, UMR 5187, 9 av. du Colonel Roche, BP 44346, 31028 Toulouse Cedex 4 \\
$^3$Tsinghua University, Beijing\\
$^4$Institute of Physics, Faculty of Philosophy and Science, Silesian University in Opava, Bezru\v{c}ovo n\'{a}m. 13,CZ-74601 Opava, Czech Republic
 }
\date{Accepted 2009 September 15.  Received 2009 September 15; in original form 2009 September 2}
\maketitle
\begin{abstract}
The width ($\Delta\nu$), root mean squared amplitude ($r_S$) of lower and upper kHz quasi-periodic oscillations (QPOs) from accreting neutron stars vary with frequency. Similarly, the QPO frequency varies with the source count rate ($S$). Hence, the significance of a QPO, scaling as $S\times r_S^2/\sqrt{\Delta\nu}$ will also depend on frequency. In addition, the significance also scales up with the square root of the integration time of the Fourier power density spectrum ($T$). In practice and in most analysis, $T$ is constrained (e.g. limited by the RXTE orbital period if the source is occulted by the earth) or minimized to increase the number of detections. Consequently, depending on the way data are considered, kHz QPOs (the lower, the upper or both kHz QPOs) may be detected only over a limited range of their frequency spans or detected predominantly at some frequencies, leading potentially to biases in the observed distributions of frequencies or frequency ratios. Although subject of much controversy, an observed clustering of QPO frequency ratios around 3/2 in Sco X-1, also seen in other sources, has been previously used as an argument supporting resonance based models of neutron star QPOs. In this paper, we measure how the statistical significance of both kHz QPOs vary with frequency. For this purpose, we consider three prototype neutron star kHz QPO sources, namely 4U1636-536, 4U0614+091 and Sco X-1, whose QPO parameters have different, though representative, dependence with frequency. As the significance of QPO detection depends on frequency, we show that in sensitivity-limited observations (as in the case of the RXTE/PCA), a simultaneous detection of both the lower and upper kHz QPOs can only be achieved over limited frequency ranges. As a result, even a uniform distribution of QPO frequencies will lead to peaks (in particular around 3/2) in the histogram of ratios of simultaneous kHz QPO frequencies. This implies that the observed clustering of ratios of twin QPO frequencies does not provide any evidence for intrinsically preferred frequency ratios in those systems, thus weakening the case for a resonance mechanism at the origin of neutron star kHz QPOs.
\end{abstract}

\section{Introduction}
\label{intro}

Despite a wealth of observational data, accumulated since their discovery in 1995, there is not yet a commonly accepted model of neutron star kHz QPOs. In recent years, a lot of efforts has been devoted to studying the distribution of the observed frequencies (or frequency ratios), in an attempt to provide insights to theoretical models. This was particularly relevant for resonance based models which predicted that both the ratio of frequencies as well as the frequencies themselves should cluster around specific values. Apparent support for these models came from the observation that the ratios of simultaneous lower and upper QPO frequencies from Sco X-1 clustered around 1.5, suggestive of a 3:2 resonance mechanism \citep{Abramowicz:2003sf}. Additional support came simultaneously from the observations of black hole high frequency QPOs associated with fixed 3/2 ratios \cite[e.g.][for a review]{Remillard:2005qp}. The clustering of frequency ratios, claimed in other neutron star systems as well \citep[for 4U1636-536]{Torok:2008zp}, remains a controversial issue since then. In particular, the significance of the clustering  presented by \citet{Abramowicz:2003sf} was challenged by \citet{Belloni:2005kc}. 

The signal to noise ratio $n_\sigma$ at which aperiodic variability such as a kHz QPO is detected in a photon counting experiment is approximately \citep{1988SSRv...46..273L,van-der-Klis:1989kn}: 
\begin{equation}
n_\sigma \propto {S^2\over B+S}r_S^2\left(T\over\Delta\nu\right)^{1/2} \sim {S}r_S^2\left(T\over\Delta\nu\right)^{1/2}
\label{eq1}
\end{equation} 
where $S$ and $B$ are source and background count rate, respectively, $r_S$ is the root mean squared amplitude of the variability expressed as a fraction of $S$, $T$ the integration time and $\Delta\nu$ the bandwidth of the variability. Hereafter, we will neglect $B$ over $S$, because for most QPO sources, $B$ (a few tens of counts/s/PCU, where PCU stands for PCA Unit, 5 at most, \citet{Bradt:1993hw}) is at least a factor of 10 lower than $S$, which is typically several hundreds of counts/s/PCU.  In equation \ref{eq1}, $r_S$  and $\Delta\nu$ vary with frequency ($\nu$) \citep{2006MNRAS.370.1140B}, hence $n_\sigma$ will depend also on $\nu$. The frequency dependence of $r_S$  and $\Delta\nu$ is a non monotonic function of  frequency and has a different shape for the lower and upper kHz QPOs.  In addition, in a given system, on average larger luminosities (hence count rates, $S$) correspond to larger $\nu$ \citep{Ford:2000ad}. Actually, the situation is more complex, because of the parallel track phenomenon, which shows that $\nu$ is not uniquely determined by $S$, but that the spread of $S$ over which QPOs are detected increases with frequency \citep{van-der-Klis:2001vl}. This is illustrated in Figure 2 in \citet{2005MNRAS.361..855B} where, in a frequency versus count rate diagram, it is mostly the left upper part of the diagram that is populated. The implications of the above are twofold. First, the integration time at which a kHz QPO is detected above a certain significance threshold will depend on frequency. Second, if $T$ is constrained, kHz QPOs, either the lower, the upper or both may reach the significance threshold only over limited frequency ranges. This directly implies that the observed distribution of frequencies or frequency ratios will strongly depend on the way the data are considered. 

When dealing with frequency distributions, two main types of analysis have been applied so far, both suffering from intrinsic biases. Methods consisting of averaging large amount of data, based on some selection criteria, e.g. spectral colors, position along a color-color diagram \citep[e.g.][for GX5-1]{jonker02}, same type of QPO falling in a given frequency interval, as for the shift and add procedure \citep[e.g.][for Sco X-1]{2000MNRAS.318..938M}, erase the underlying distribution of frequencies and cannot be considered for this purpose. The first applicable method considers observations of similar durations. This is the case in particular when data are analyzed ObsID per ObsID, whose durations are limited by the RXTE orbital period to typically $\sim 2-3$ kseconds (by definition, an ObsID corresponds to a single observation, where observation refers to a temporally contiguous collection of data from a single pointing). This method will obviously bias the distribution of frequencies or ratios at frequencies where the two QPOs can be detected simultaneously on a typical ObsID timescale. Such a method was recently applied to 4U1636-536 by \citet{Torok:2008zp} who reported a clustering of ratios at 3/2 and possibly at 5/4, to 4U1820-303 by \citet{2008NewAR..51..835B} who found a clustering of ratios of simultaneous twin QPOs around 4/3 and to 4U0614+091 by \citet{2009arXiv0907.3223B}. In both studies, it was found that the distribution of ratios computed from the single lower QPO frequencies, estimating the upper QPO frequencies from their linear relationship differed from the distribution of ratios of simultaneous twin QPOs, as expected if the condition for detecting the lower QPO differed from the one for detecting the two QPOs simultaneously. 

As to increase the number of detected frequencies, the second method minimizes the PDS integration time to detect one of the two QPOs (generally the lower QPO) and computes the missing frequency through the linear relationship linking the lower and upper QPO frequencies. This method will bias the distributions at frequencies where the lower QPO is easy to detect (e.g. when its RMS amplitude/quality factor reaches a maximum). It was used by \citet{Belloni:2005vl} who averaged a variable number of PDS (with integration time ranging from 64 seconds to 1280 seconds) to obtain a significant detection of the lower QPO. They found that the distribution of lower QPO frequencies showed a pronounced peak around 850 Hz in 4U1636-536. Other less prominent peaks were present in the distribution of frequencies, all centered on different ratios. Later on, \citet{Belloni:2007ad} monitored 4U1636-536 for 1.5 years, during regular 2 kilosecond pointings separated by 2 days, and found that the distribution of lower QPO frequencies differed from the previous one. In particular, the frequency distribution of the lower QPO extended down to 600 Hz and the overall distribution was flatter than previously measured. They also noticed that when considering one frequency per segment of 2 kseconds, the upper kHz QPO was detected only between 520 Hz and 950 Hz with an even distribution. All this led \citet{Belloni:2007ad} to conclude that the kHz QPOs in 4U1636-536 did not show intrinsically preferred frequencies.

\begin{figure*}
   \begin{center}
   \includegraphics[width=.485\textwidth]{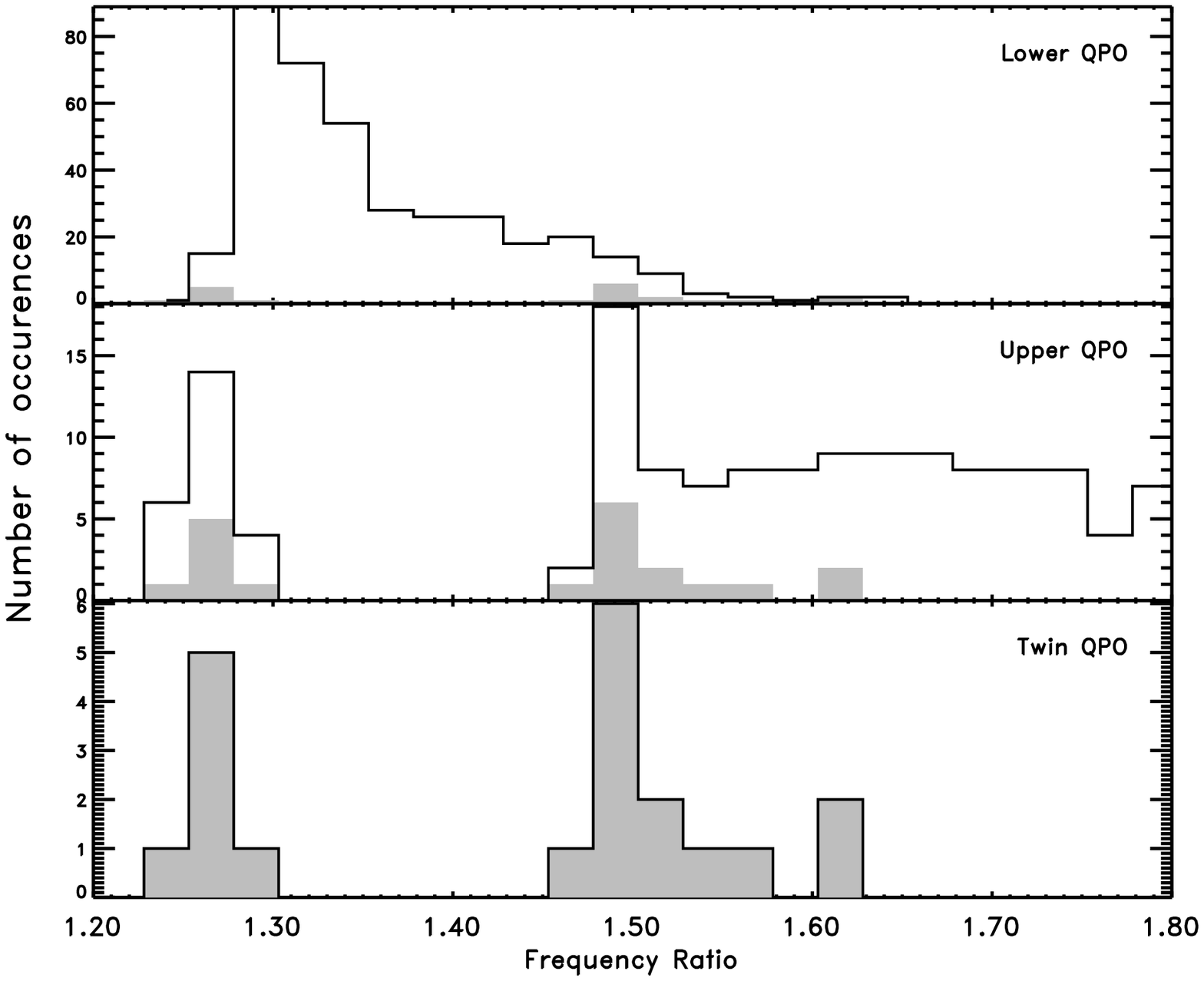}\includegraphics[width=.485\textwidth]{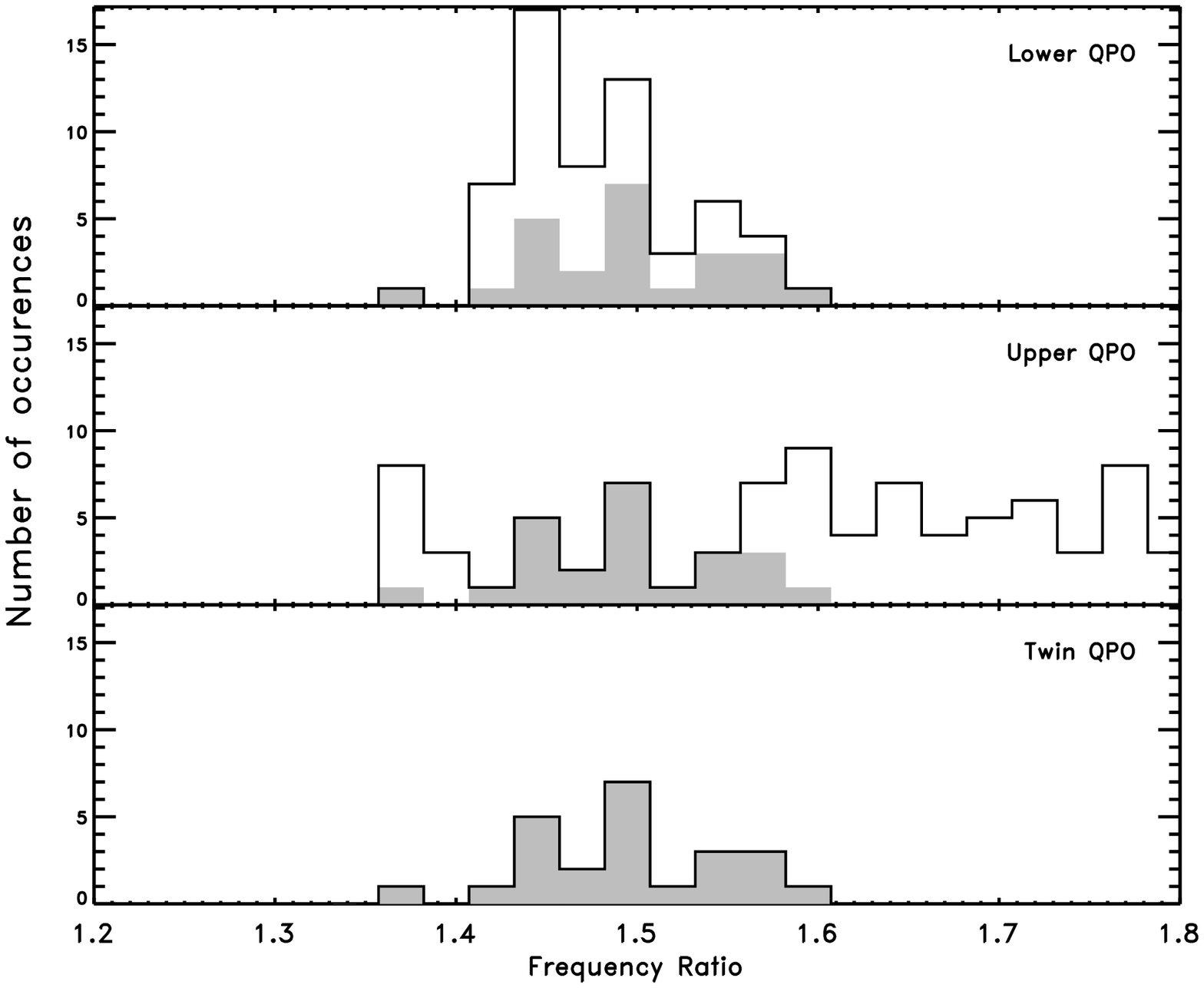}
      \caption{Histograms of ratios estimated from the frequencies of all single lower QPOs, all lower in twin QPOs (area in grey), all single upper QPOs, all upper in twin QPOs (area in grey) and all twin QPOs. In the case of a single QPO, the frequency of the other was computed from the linear function linking the two frequencies. 4U1636-536 is shown on the left and 4U0614+091 on the right hand side. Each frequency in that plot corresponds roughly to a similar integration time of the PDS (a typical ObsID duration).}
   \label{boutelier:fig1}
   \end{center}
\end{figure*}

In this paper, we measure the frequency dependence of $n_\sigma$ and show how it affects the distribution of frequencies measured with the two methods discussed above. This paper extends on previous work by \citet{Torok:2008lr} who did not directly compute $n_\sigma$ but rather scaled its value as the product of  $r_S^2/\sqrt{\Delta\nu}$: the frequency dependence of $r_S$ and $\Delta\nu$ was recovered for 4U1636-536 by interpolating the values reported by \citet{2005MNRAS.361..855B}, i.e. derived from the analysis of a subset of the data considered here. Here, we have chosen three systems 4U1636-536, 4U0614+091 and Sco X-1, as being representative of low and high luminosity neutron star QPO sources, both in terms of RMS amplitudes and quality factors. Both sources have been extensively observed by RXTE; several hundreds of ObsIDs spanning over more than 10 years, thus minimizing any biases in the frequency distribution of their QPOs that could arise as a result of a non uniform sampling of the source states. This matters because \citet{Belloni:2005vl} have shown that a time limited and sparse sampling of a source, whose QPO frequency was modeled by a random walk could produce peaked frequency distributions. In addition, as discussed by \citet{Belloni:2007ad}, in the case of 4U1636-536, the long term evolution of the source intensity may also affect the observed frequency distributions.

In the following section, we present the histogram of frequency ratios, as derived from the homogenous processing of all archival data available to date for both 4U1636-536 and 4U0614+091. We then describe the analysis scheme used to recover $n_\sigma$ over the full frequency span of both the lower and upper QPOs.  Then, knowing $n_\sigma$ as a function of frequency, assuming an uniform distribution of frequencies for the upper QPO, we show that we can reproduce, without any further assumptions, the clustering of frequency ratios seen in the data. We also illustrate how the method of optimizing the PDS integration time introduce strong biases in the histogram of ratios. Finally, we then extend our analysis to Sco X-1 before concluding.

\section{Data analysis}
We have first considered all archival data from the Rossi X-ray Timing Explorer for 4U1636-536 and 4U0614+091. Our analysis builds upon previous work by \citet{2005MNRAS.361..855B, 2005AN....326..808B, 2006MNRAS.370.1140B, 2007MNRAS.376.1139B} for 4U1636-536 and by \citet{2009arXiv0907.3223B} for 4U0614+091. Data are considered per ObsID. For each ObsID, we have computed 16 second PDS with a 1 Hz resolution, using events recorded between 2 and 40 keV. The PDS are normalized according to \citet{Leahy:1983mb}, so that the Poisson noise level is constant around 2.  As previously shown for 4U1636-536, the lower QPO can be so strong and so narrow that in some ObsIDs, the frequency drift within the ObsID can be corrected for using a tracking procedure similar to the one described in \citet{2005MNRAS.361..855B}.  Whenever possible, all the 16 second PDS are then aligned to the mean frequency of the lower QPO within the ObsID, to produce one single ObsID averaged PDS (otherwise they are averaged directly). This enables us to obtain meaningful parameters (width and amplitude) averaged over the ObsID, helping us to detect the upper QPO in some case, and in all cases, to identify wether it is a lower or an upper kHz QPO (see below). In 4U0614+091, the QPO is not strong enough to enable such a analysis. The ObsID averaged PDS is then blindly searched for excess power between 300 Hz and 1400 Hz using a scanning technique, as presented in \citet{Boirin:2000jt}. Each excess (at most the 2 strongest) is then fitted with a Lorentzian with three free parameters; frequency, full width at half maximum (constrained to range from 2 to 1000 Hz), and amplitude (equal to the integrated power of the Lorentzian). The Poisson noise level is fitted separately above 1400 Hz and then frozen when fitting the QPOs (the fitted noise level is generally close to 2, indicating no significant deadtime). Errors on each parameter are computed with $\Delta\chi^2=1$. As in previous papers in this field, our threshold for QPOs is related to the ratio (hereafter $R$) of the Lorentzian amplitude to its $1\sigma$ error ($R$ was often quoted and used as a significance).  In this paper, our threshold for $R$ is 3, meaning that we consider only QPOs for which we can measure the power of the Lorentzian with an accuracy of $3\sigma$ or more. Such a  threshold corresponds to a  $\sim 5.5-6\sigma$ excess power in the PDS for a single trial, equivalent to $\sim 4\sigma$ significance if we account for the number of trials of the scanning procedure \citep{van-der-Klis:1989kn}. 
\subsection{Histograms of ratios}
At the end of this first stage of our analysis, we have a set of ObsID averaged PDS (i.e. of comparable durations) with one or two significant QPOs in. For those averaged PDS in which only one significant QPO is detected, we identify whether it is a lower or an upper by placing the QPO parameters in a RMS--$\nu$ and quality factor--$\nu$ diagram following \citet{2006MNRAS.370.1140B}. In addition to the histogram of ratios derived from simultaneous twin QPO frequencies, using the nearly linear relationship between the lower and upper QPO frequencies, it is possible to compute the histograms of ratios estimated from the frequencies of all single lower QPOs and all single upper QPOs. The histograms are presented in Figure \ref{boutelier:fig1}; the frequencies entering the histograms are the mean frequencies over the ObsID. The clusters previously found by \citet{Torok:2008zp} around 1.5 and 1.25 for 4U1636-536 are reproduced and clearly the histograms computed from either the single lower or single upper QPOs is different from the one computed from simultaneous twin QPOs. A cluster of simultaneous twin QPOs centered around 1.5 is also seen from 4U0614+091. 
\subsection{Frequency dependence of QPO significance}
We now wish to estimate how the significance of kHz QPOs varies with frequency. For this purpose, we apply the shift-and-add technique \citep{Mendez:1998qf}. As originally discussed, this technique enables us to combine large amount of data, so that both the lower and upper QPOs can be detected simultaneously (this is possible for some ObsID, but generally requires longer integration times of the PDS). We shift-and-add all ObsID averaged PDS, considering all those with a QPO of the same type falling in the same frequency interval. The width of the frequency interval has been chosen to group enough PDS as to detect the two QPOs simultaneously, with a high significance. The large amount of data in the RXTE archive makes possible to consider interval of 30 Hz for 4U1636-536 and 50 Hz for 4U0614+091. Obviously, not all frequency intervals correspond to the same total PDS integration time, so we normalized the significance using equation \ref{eq1} to the square root of the observing time of a typical ObsID, which we assumed to be 3 kseconds (this is about the mean duration of all ObsIDs from 4U0614+091: it is closer to 2.5 kseconds for 4U1636-536).  
\begin{figure*}
   \begin{center}
   \includegraphics[width=.485\textwidth]{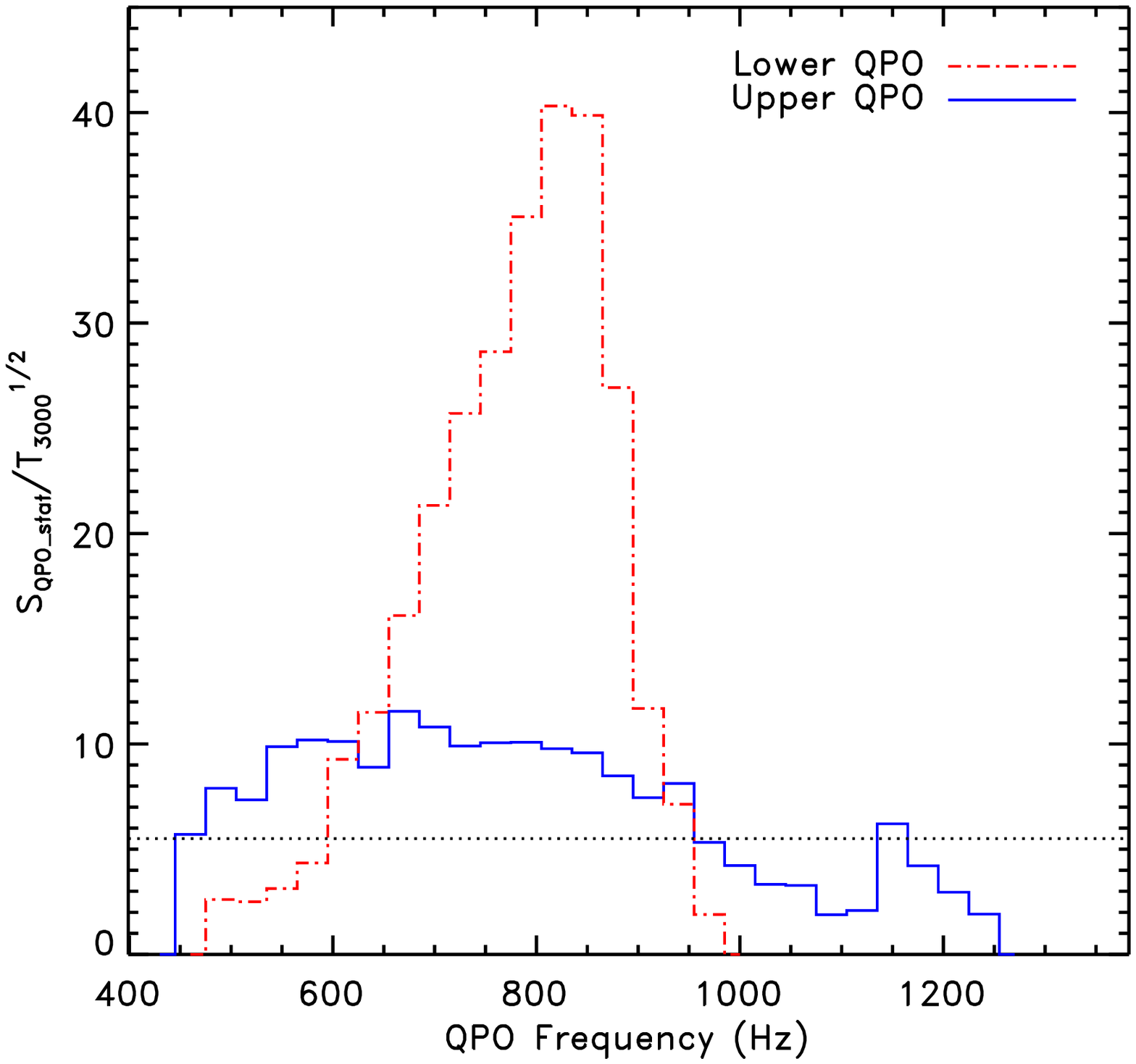}\includegraphics[width=.485\textwidth]{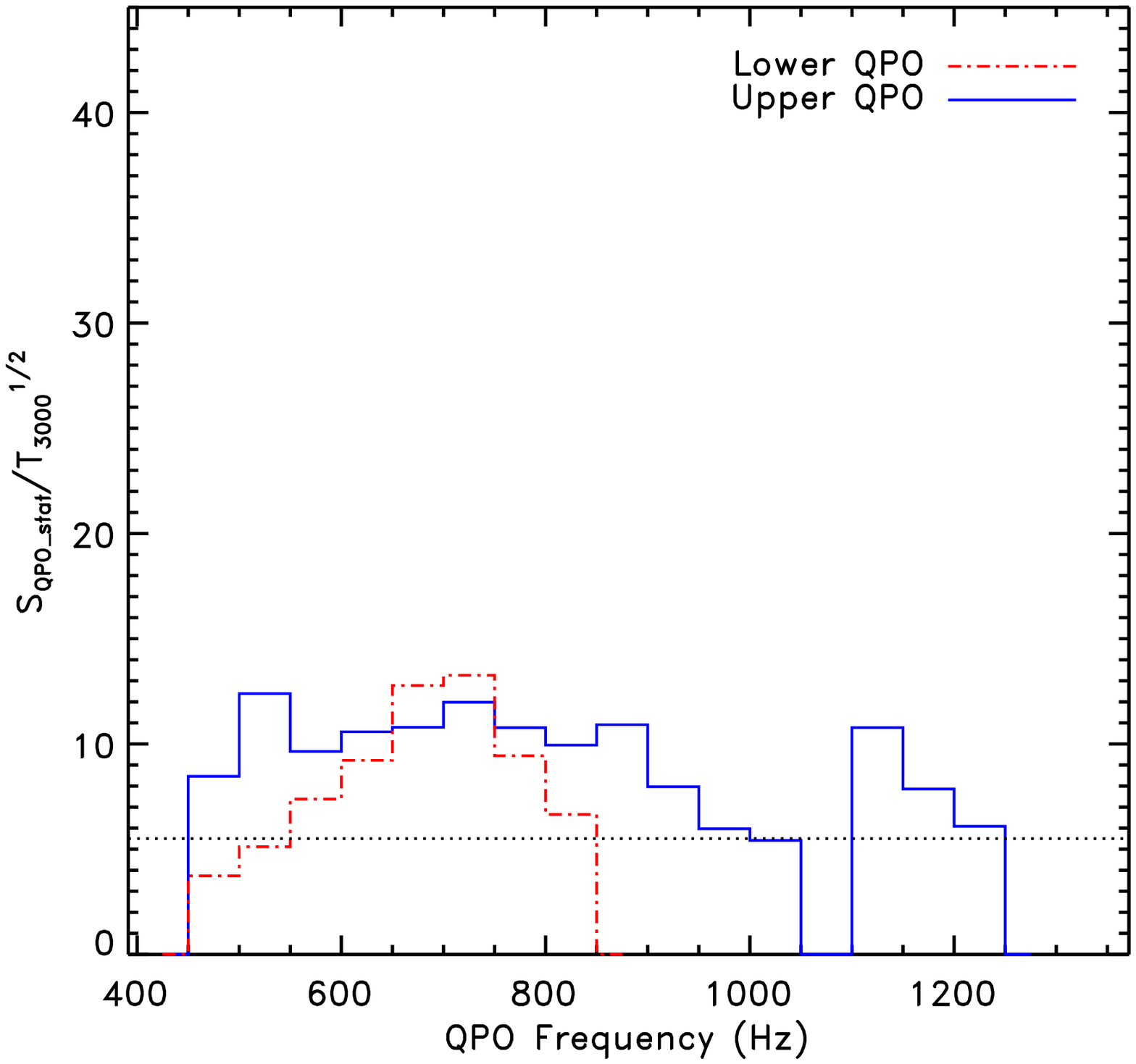}
      \caption{The significance of the PDS excess power in 4U1636-536 (left) and 4U0614+091 (right) against frequency, as recovered after applying the shift-and-add technique and normalizing the significance to the same observing time, assumed to be 3 kseconds, i.e. comparable to a typical ObsID exposure. The significance is compute from \citet{Boirin:2000jt}. It is equivalent to the significance given in equation 1 with a 1/4 scaling coefficient. The horizontal dashed line indicates the 5.5 sigma (single trial) statistical significance threshold used in our analysis. The same way that not all frequency bins correspond to the same PDS integration time, not all frequency bins correspond to the same number of active PCUs. To account for this, the significance has also been normalized to the mean number of active PCUs over the whole frequency span (typically around 4). Neither the lower or upper QPO can be detected over their full frequency span on such a timescale, considering a threshold of 5.5$\sigma$. In particular, the gap in the distribution of frequencies of the upper QPO visible in 4U1636-536 \citep{2006MNRAS.370.1140B} and in 4U0614+091 \citep{2009arXiv0907.3223B} is now explained, as being caused by the significance being lower than the threshold generally used in previous analysis. We use the same scale of the y axis to show that the lower QPO is much easier to detect in 4U1636-536 than in 4U0614+091. }
   \label{boutelier:fig2}
   \end{center}
\end{figure*}

The result of such analysis is shown on Figure \ref{boutelier:fig2} for both 4U1636-536 and 4U0614+091, which shows the frequency dependence of the QPO significance (excess power in the PDS, single trial) normalized to the same integration time. Figure \ref{boutelier:fig2} demonstrates that neither the lower or upper QPO are detected over their full frequency span, for a commonly used single trial significance threshold of $5.5\sigma$. 
\begin{figure*}
   \begin{center}
   \includegraphics[width=.485\textwidth]{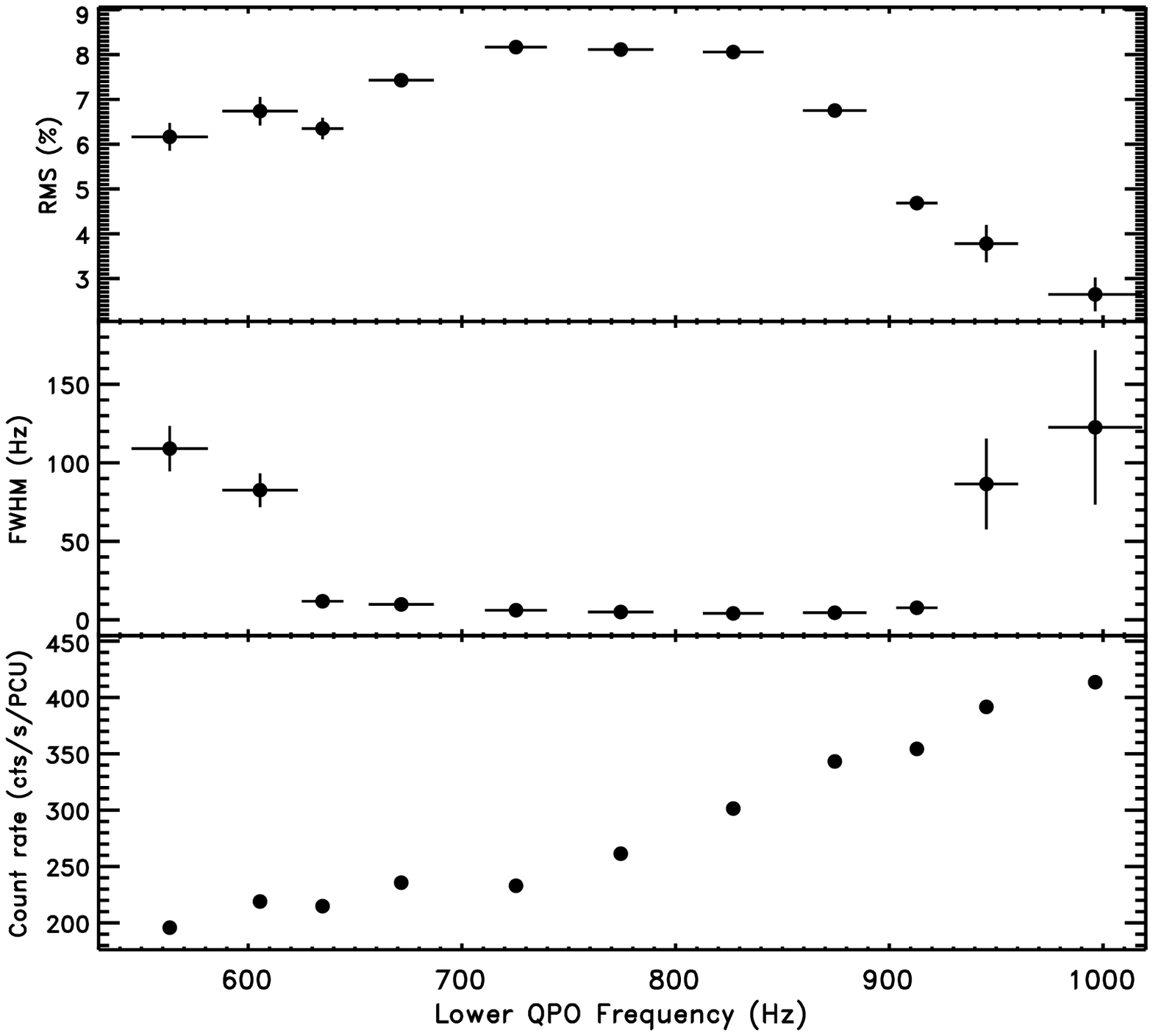}\includegraphics[width=.485\textwidth]{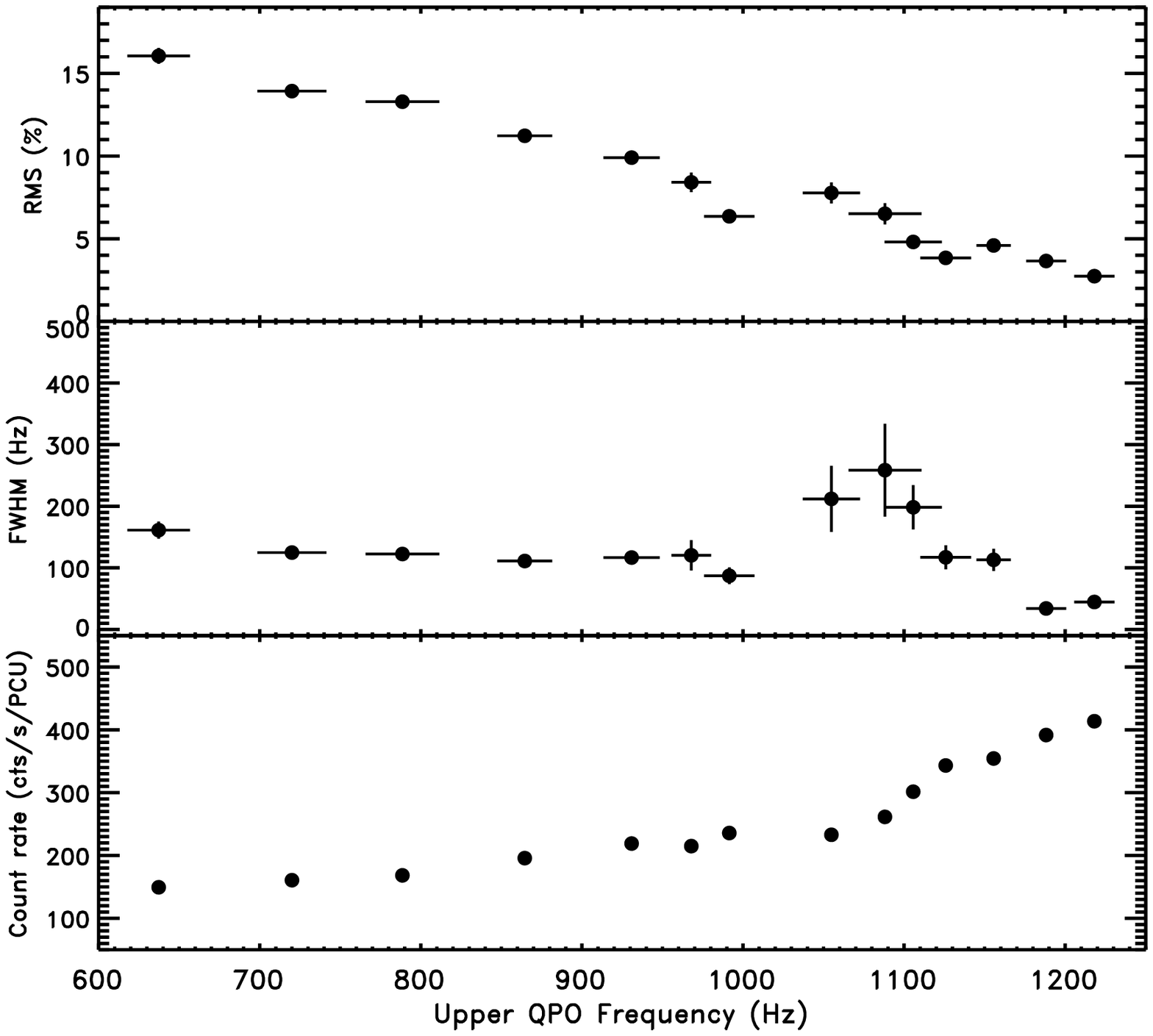}
      \caption{The RMS amplitude ($r_S$), the width ($\Delta\nu$), and the mean source count rate (normalized by the mean number of operating PCUs) for the lower  (left) and upper (right) QPO in 4U1636-536, as averaged over frequency bins of 50 Hz and 75 Hz respectively. This is the combination of these three parameters as described in equation 1 which explains the dependence of $n_\sigma$ with frequency shown in Figure \ref{boutelier:fig2}. The local minimum of $n_\sigma$ around 1100 Hz visible in Figure \ref{boutelier:fig2} for the upper QPO is related to a local increase of the QPO width.}
   \label{boutelier:fig3}
   \end{center}
\end{figure*}
\begin{figure*}
   \begin{center}
   \includegraphics[width=.485\textwidth]{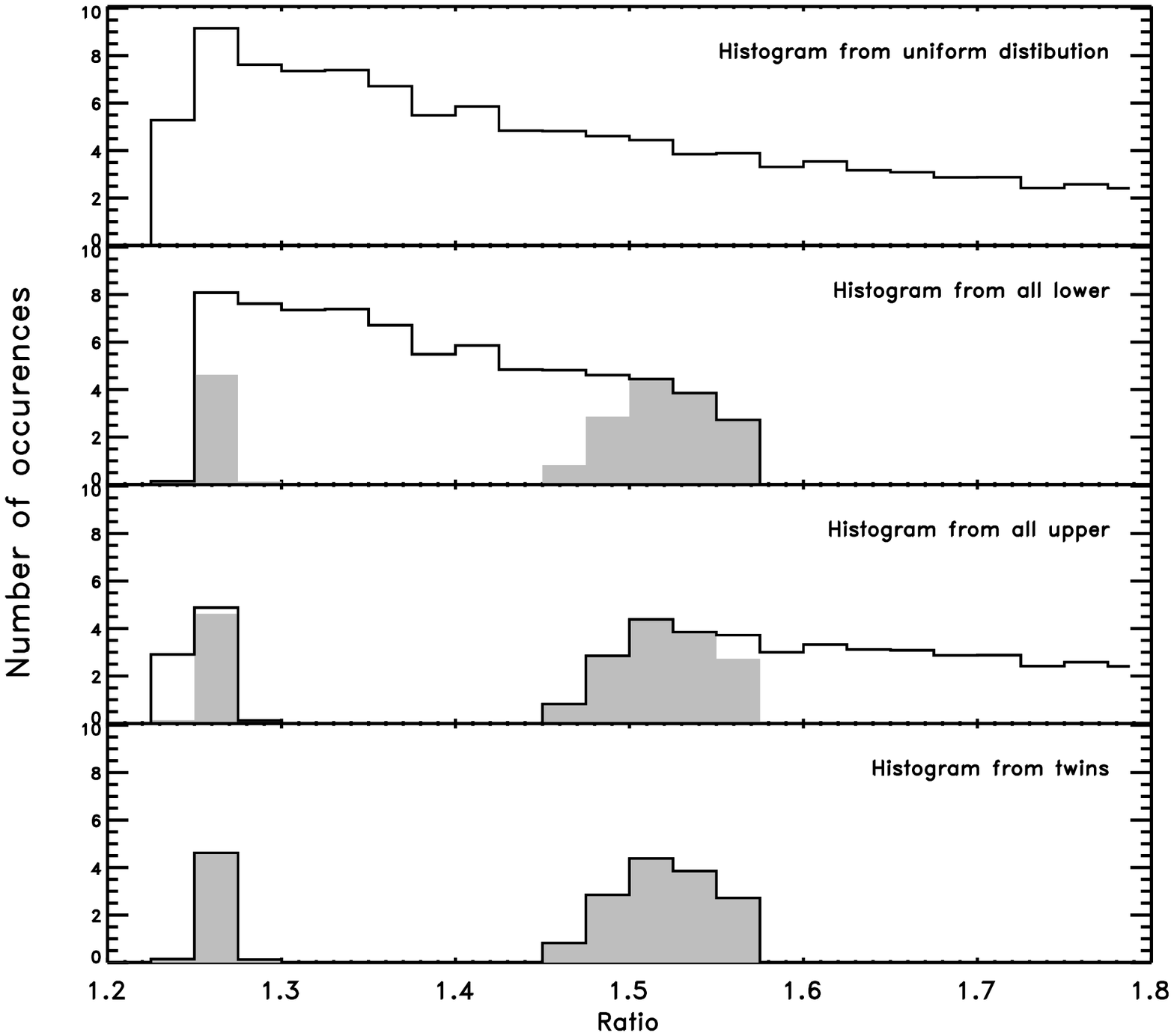}\includegraphics[width=.485\textwidth]{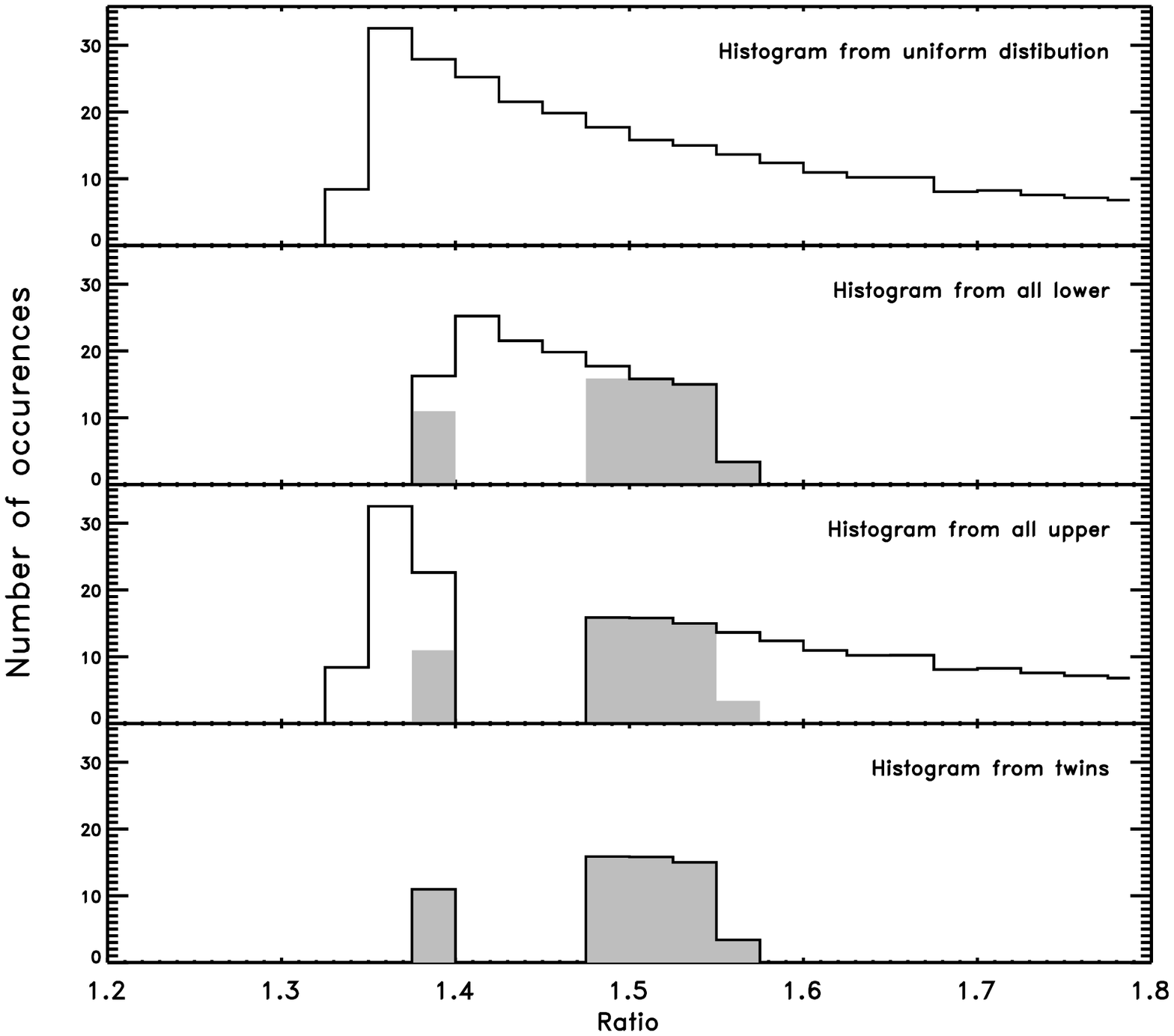}
      \caption{Simulated histograms of ratios expected from a uniform distribution (top panel, assuming that the upper QPO frequency distribution is uniform between 550 Hz and 1250 Hz), and the one derived from the frequencies of all single lower QPOs (second panel from top, the shaded area recalls the histogram of ratio from simultaneous twin QPO frequencies shown at the bottom), all single upper QPOs (third panel from top, the shaded area recalls the histogram of ratio from simultaneous twin QPO frequencies), and all twin QPOs (bottom panel). As first pointed out by \citet{Belloni:2005vl}, as a consequence of the nearly linear relationship between the two QPO frequencies, note that the histogram of ratios from a uniform distribution of frequencies is peaked towards lower ratios. In the case of a single QPO, the frequency of the other was computed from the linear function linking the two frequencies. 4U1636-536 and 4U0614+091 are shown on the left and right hand side respectively. The main peaks as well as the secondary peaks of the observed distributions are nicely reproduced with the simulations, even with our simplified assumptions.}
   \label{boutelier:fig4}
   \end{center}
\end{figure*}
As discussed earlier,  $n_\sigma$ depends on $S$, $r_S$ and $\Delta\nu$, how those quantity vary with frequency for the lower kHz QPO in 4U1636-536 is shown in 
Fig. \ref{boutelier:fig3}. The maximum of $n_\sigma$ corresponds to the maximum of $r_S$ and the minimum of $\Delta\nu$ (or the maximum of the QPO coherence). From the lowest QPO frequency to the highest one, the mean count rate increases by less than a factor of 2.

For the upper QPO, as shown in Figure \ref{boutelier:fig3}, $n_\sigma$ is rather flat: the decrease of $r_S$ is compensated by a decrease of $\Delta\nu$ and an increase of $S$. There is a local minimum around 1100 Hz in both sources. For 4U0614+091, \citet{2009arXiv0907.3223B} have shown that this was related to a local decrease of $r_S$. For 4U1636-536 (and also for 4U1820-303, \citet{2008NewAR..51..835B}), it is associated with a local increase of the width of the upper QPO (see Fig. \ref{boutelier:fig3}). This local minimum provides an explanation for the gap in the distribution of frequencies of the upper QPO visible in 4U1636-536 \citep{2006MNRAS.370.1140B} and in 4U0614+091 \citep{2009arXiv0907.3223B} (when data are analyzed per ObsID). After the local minimum, $n_\sigma$ reaches a local maximum, enabling us to detect twin QPOs over a narrow range of frequencies, which in turn produces a second cluster of ratios, e.g. below 1.3 in 4U1636-536.

\section{Simulations}
Now we wish to visualize the influence of the frequency dependence of the significance of the QPO on the observed distribution of ratios. For this, we assume an uniform distribution of frequencies for the upper QPO which spans a wider range of frequencies than the lower QPO. For example, for 4U1636-536, the upper QPO frequency spans from 550 Hz to 1250 Hz, while the lower QPO frequency varies only from 550 Hz to 950 Hz. Obviously, by drawing a uniform distribution of frequencies within the range of the lower QPO frequency, one would miss all the single upper QPO frequencies below $\sim 800$ Hz (those QPOs populate the right part of the histograms of ratios predicted from the upper QPO also, middle panel in Figure \ref{boutelier:fig1}). For each upper QPO frequency so generated, we compute the frequency of the lower QPO, from a linear function $\rm \nu_{upper}=a \times \nu_{lower}+b$ with $\rm a=0.70$ and $\rm b=520$ for 4U1636-536 and $\rm a=1.0$, and $\rm b=320$ for 4U0614+091 respectively. The a and b parameters were derived for each source by fitting twin QPO frequencies detected simultaneously over the same ObsID. We have checked that the parameters are consistent with each other when fitting the twin QPO frequencies recovered from the shift-and-add procedure described above. Instead of fitting with a linear function, we have also considered a power law function, leading to very similar results, as expected since the two functions are essentially identical in the range where the two QPOs can be detected simultaneously. We discard those predicted lower QPO frequencies falling outside the range of detected frequencies, i.e. below 500 Hz and above 950 Hz. We interpolate between the values of n$_\sigma$ shown in Figure \ref{boutelier:fig2} to estimate the corresponding significance for all generated lower and upper QPO frequencies. 
\subsection{Histograms with a constant integration time}
We first consider the case in which the significance of each QPO is computed for a similar integration time (3 kseconds), i.e. reproducing the case where the real data are analyzed ObsID per ObsID (the frequency generated represents the mean QPO frequency within an ObsID). The results of this simulation, considering a significance threshold of $5.5\sigma$, are shown in Figure \ref{boutelier:fig4} in a form similar to Figure \ref{boutelier:fig2}. What is striking from this figure is that, even this oversimplified simulation, reproduces the shape of the histograms and the clustering of ratios seen in the data. For example, the second cluster of ratios of twin QPOs in 4U1636-536 below 1.3 corresponds to the local maximum of the $n_\sigma$ curve for the upper QPO around 1150 Hz. Finally, it is obvious that the distributions of ratios computed from one of the two QPO frequencies are also not complete and not representative of the underlying distribution. The good agreement between the simulations and the data clearly indicates that the observations performed by RXTE over the $\sim 10$ year period have provided a close to uniform sampling of the kHz QPOs over their entire frequency span.
\subsection{Histograms with a variable integration time}
\begin{figure}
   \begin{center}
   \includegraphics[width=.485\textwidth]{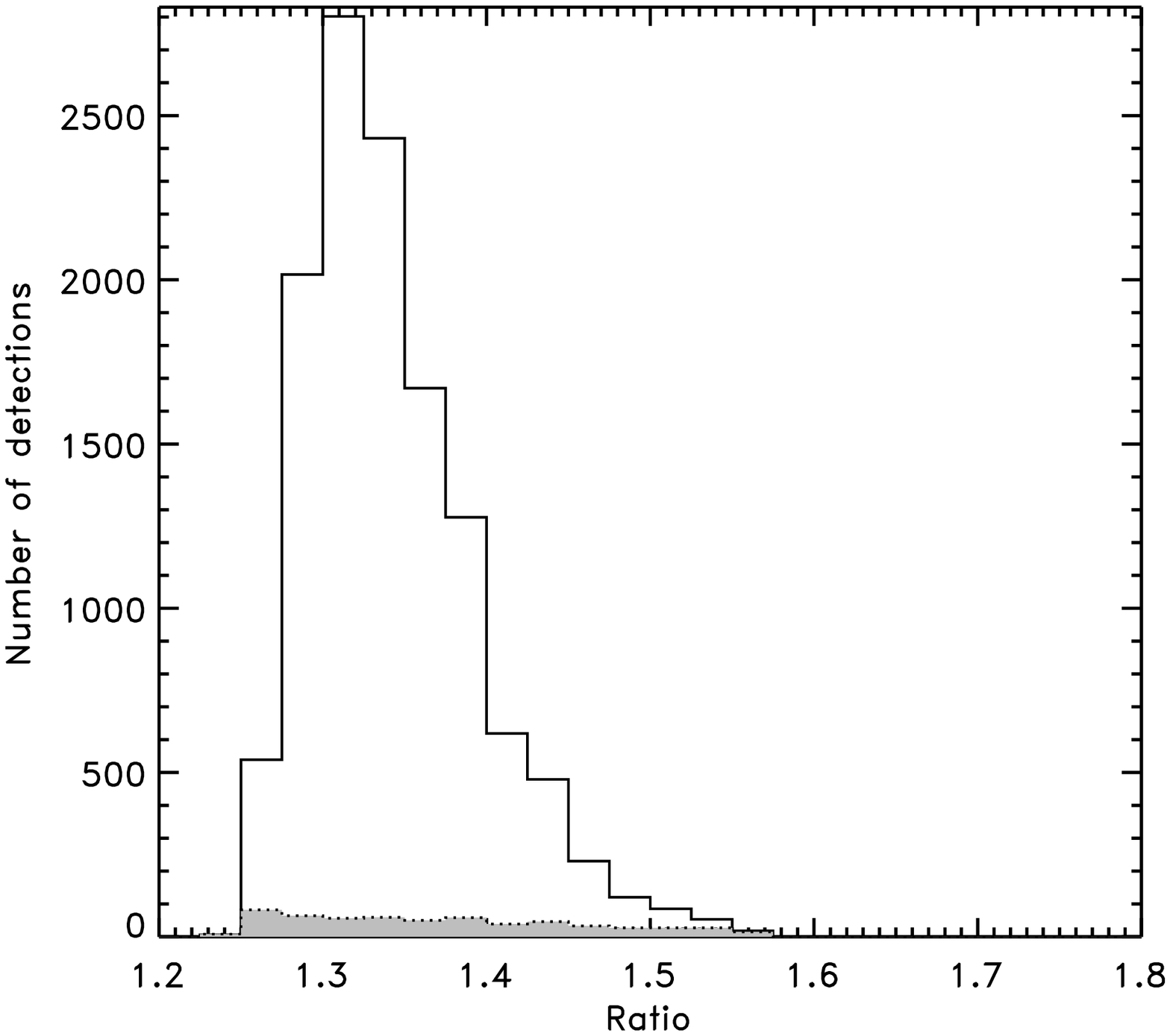}
      \caption{Simulated histogram of ratios estimated from the frequencies of all single lower QPOs, assuming a uniform distribution of frequencies, and a variable integration time of the PDS for detecting the QPO. The histograms peaks at the ratio which corresponds to a frequency of $\sim 850$ Hz for the lower QPO, i.e. where it can be followed on the shorter timescales. The shaded region delimited by a dashed line corresponds to the ratio histogram, expected from the uniform distribution, and considering one frequency per observation.}
   \label{boutelier:fig5}
   \end{center}
\end{figure}

We have simulated again a uniform distribution of lower QPO frequencies, generating 590 random frequencies corresponding to 590 ObsIDs of 3 kseconds (as is the case for 4U1636-536). Using Figure \ref{boutelier:fig2} and equation 1, it is possible to compute the minimum PDS integration time, required for the QPO to exceed a significance threshold of $5.5\sigma$, hence the number of detections associated with each segment of 3 kseconds (e.g. for 4U1636-536, at 850 Hz, the significance reaches $\sim 40\sigma$ for 3 kseconds: this corresponds to $5.5\sigma$ in about 60 seconds, hence a segment of 3 kseconds will contribute 50 frequencies around 850 Hz in the histogram). Having the lower QPO frequencies, the upper QPO frequencies are estimated with the linear function described above. From this, one can produce an histogram with all the detections, keeping in mind that not all detections correspond to the same PDS integration time. An example of such simulation is shown in Fig \ref{boutelier:fig5}. The histogram of ratios is strongly peaked around 1.3, as found in data analyzed by the same method by \citet{Belloni:2005vl}: the main peak was found at 1.30$\pm 0.02$ for $\nu=856 \pm 2 $. Such histogram (which just reflects the variation of $n_\sigma$ with frequency) is much more biased than the histogram of single lower QPO computed from detections over similar integration times (see Figure \ref{boutelier:fig4}, left hand side, second panel from the top). 
\subsection{More realistic simulations}
\begin{figure}
   \begin{center}
   \includegraphics[width=.485\textwidth]{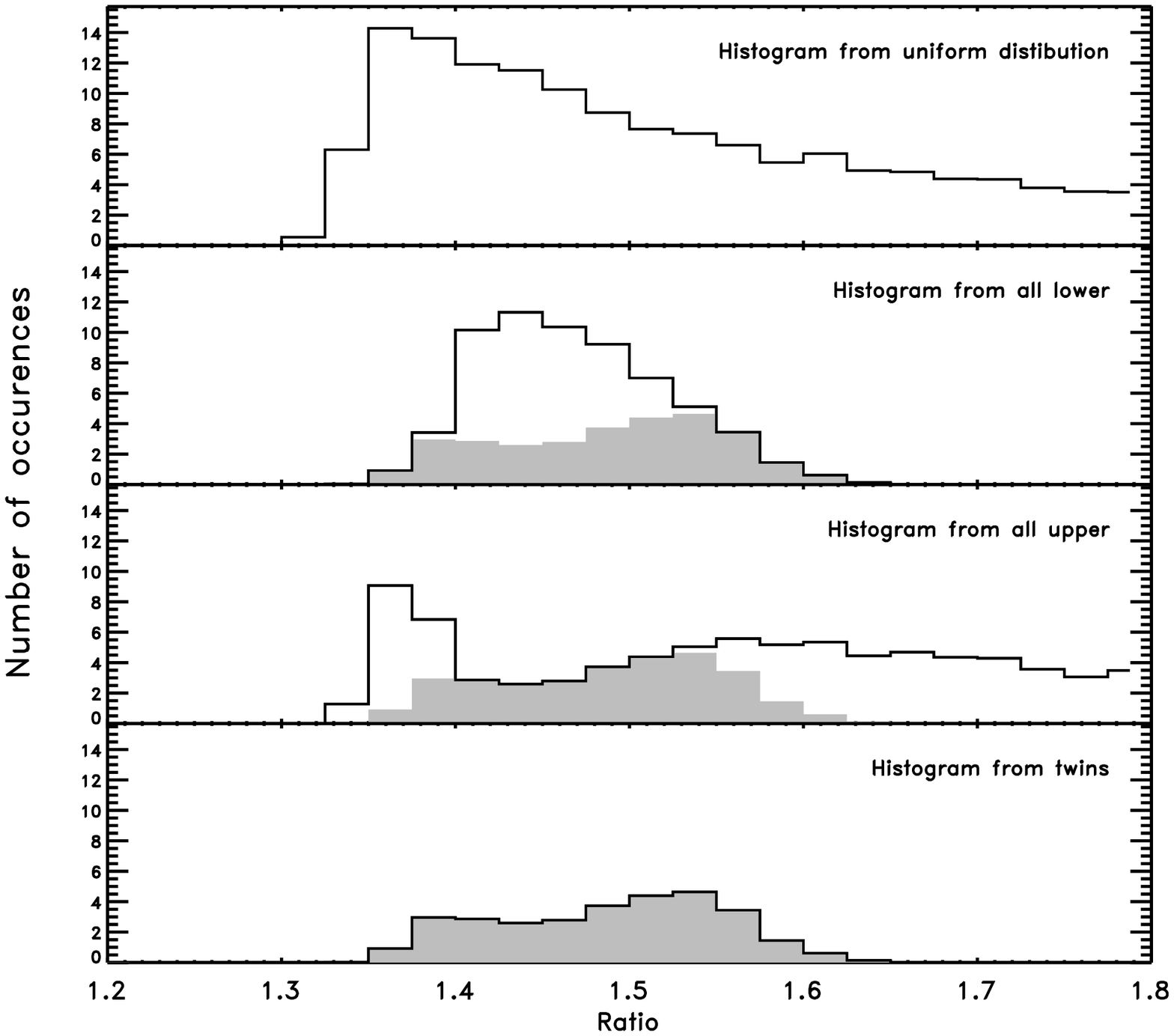}
      \caption{Simulated histogram of ratios for 4U0614+091 for a more sophisticated simulation, which accounts for the fact that the spread of count rates at which QPOs are detected increases with frequencies (the parallel tracks), for a spread in the exposure time between 2 and 4 kseconds, for a spread in the number of active PCUs (between 3 and 5) and finally for a 10 Hz jitter in the relation linking the two QPO frequencies. As a result, the simulated clusters show a wider spread more in line with the real data (see Figure \ref{boutelier:fig1}.)}
   \label{boutelier:fig6}
   \end{center}
\end{figure}

We have also added complexity to our simulations by modeling the parallel-track phenomenon (or the frequency dependence with count rates), the spread of the ObsID exposure times around their mean value and a varying number of active PCA units (the latter two quantities being obviously not frequency dependent). We have also included a jitter in frequency along the linear correlation between the two frequencies. In all cases, this does not change significantly the histograms simulated, it just increases slightly the spread of the clusters, making them more similar with the data. This is illustrated for the case of 4U0614+091 in Figure \ref{boutelier:fig6}. The gap between 1.4 and 1.5 in the ratio distribution is now filled in.
\begin{figure*}
   \begin{center}
   \includegraphics[width=.485\textwidth]{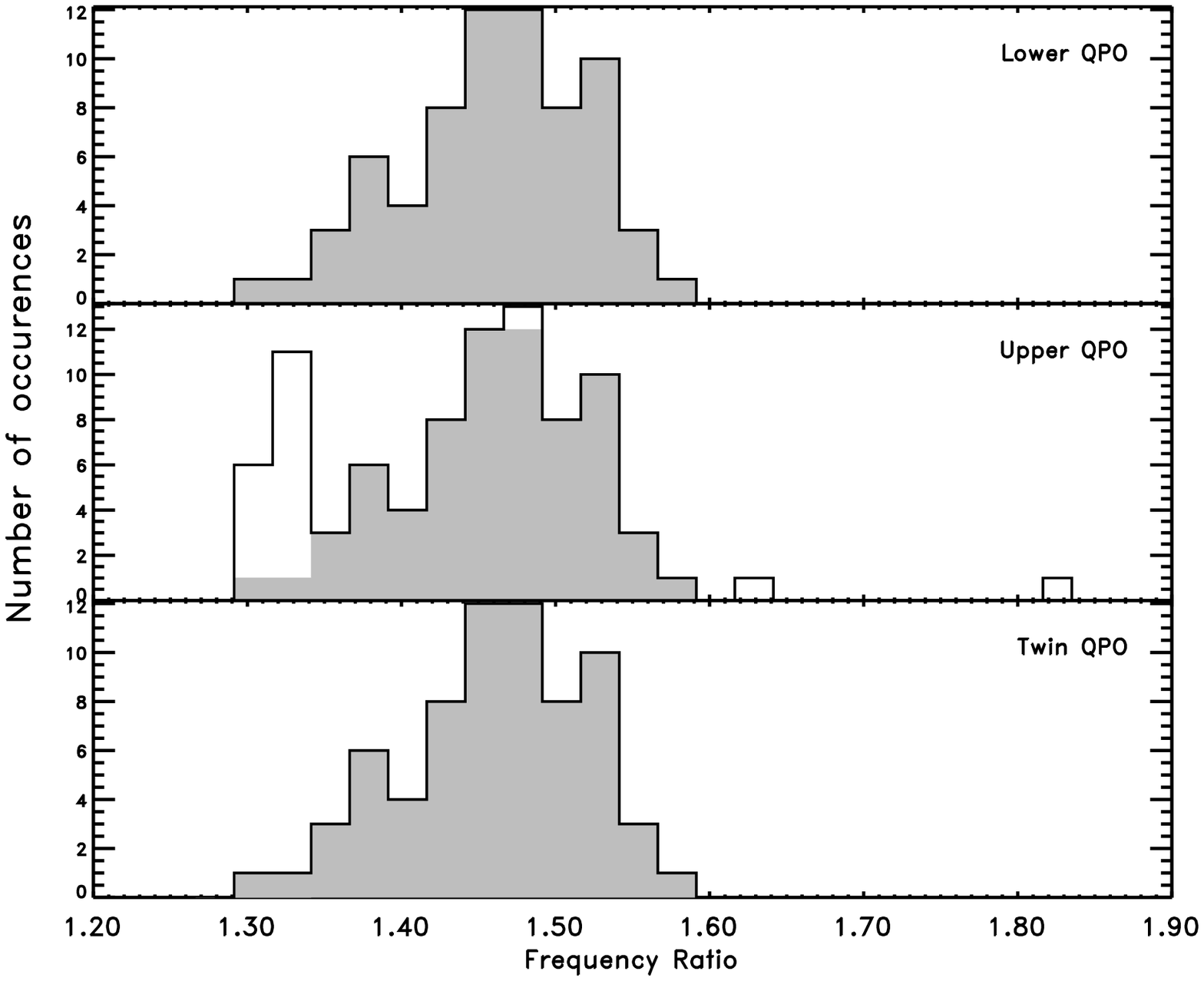}\includegraphics[width=.465\textwidth]{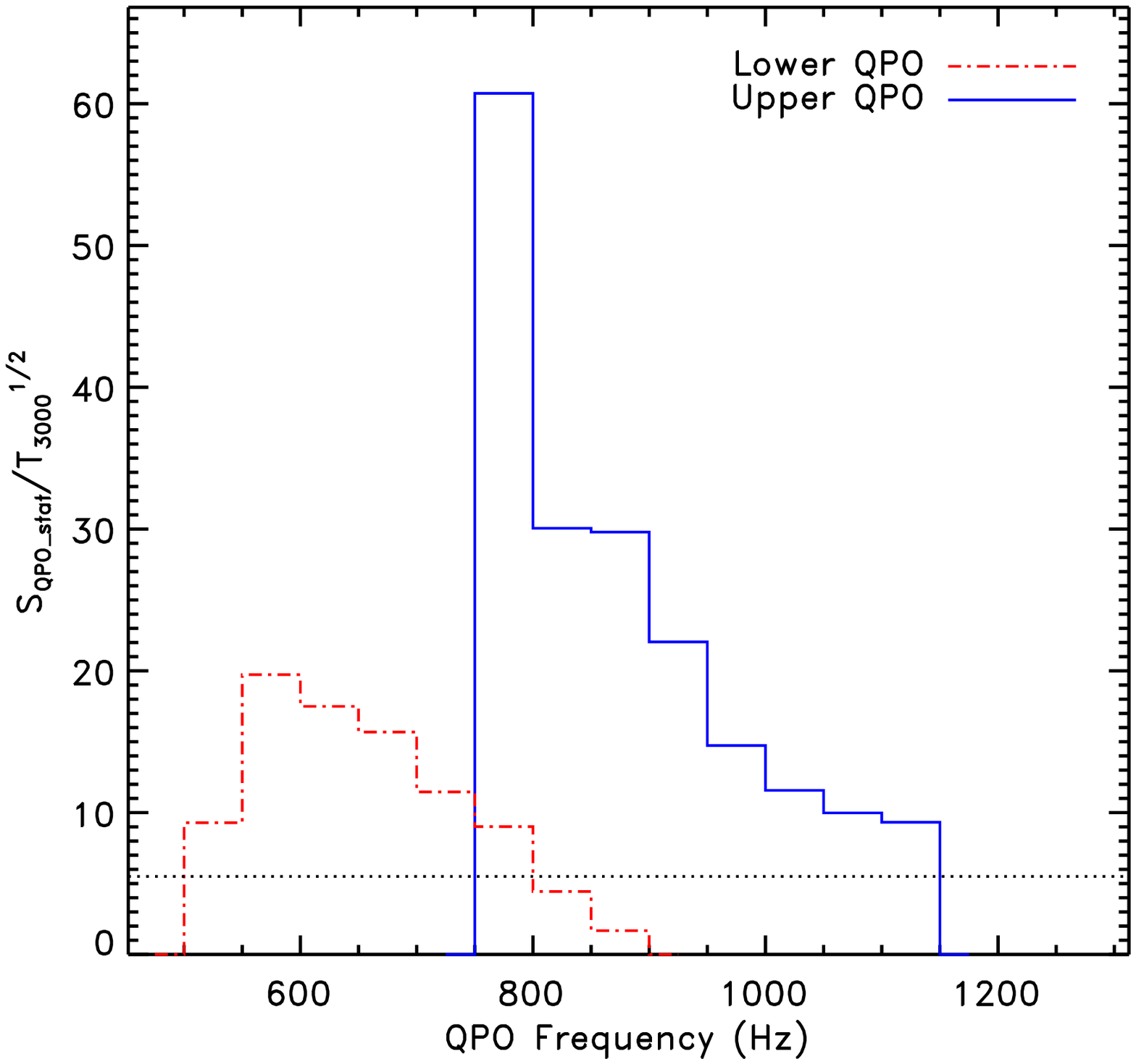}
      \caption{Left) The histogram of ratios from all lower, all upper and all twin kHz QPO frequencies. In the top two panels, the histogram in grey is the one derived from simultaneous twins. As can be seen, twin QPOs are generally detected simultaneously for Sco X-1, and there are no occurrence where the lower kHz QPO is detected alone (unlike in the two other sources). In the middle panel, note that there is one single detection of a single upper QPO, at a frequency just above $\sim 750$ Hz, corresponding to a ratio $\sim 1.8$ (this is the lowest upper QPO frequency reported for the source so far). The gap in the ratio distribution corresponds to a frequency gap about 100 Hz. Right) The significance of the PDS excess power in Sco X-1 against frequency, as recovered after applying the shift-and-add technique over frequency intervals of 50 Hz and normalizing the significance to the same observing time, assumed to be 3 kseconds, i.e. comparable to a typical ObsID exposure. The lower QPO significance is always lower than the upper QPO significance. The lower QPO is detected over a relatively narrow range of frequencies about $\sim 300$ Hz wide. The first bin of the upper QPO histogram, where the significance reaches a maximum, contains only one ObsID (the upper QPO frequency is at about 750 Hz)}
   \label{boutelier:fig7}
   \end{center}
\end{figure*}
\section{Sco X-1}
\begin{figure*}
   \begin{center}
   \includegraphics[width=.485\textwidth]{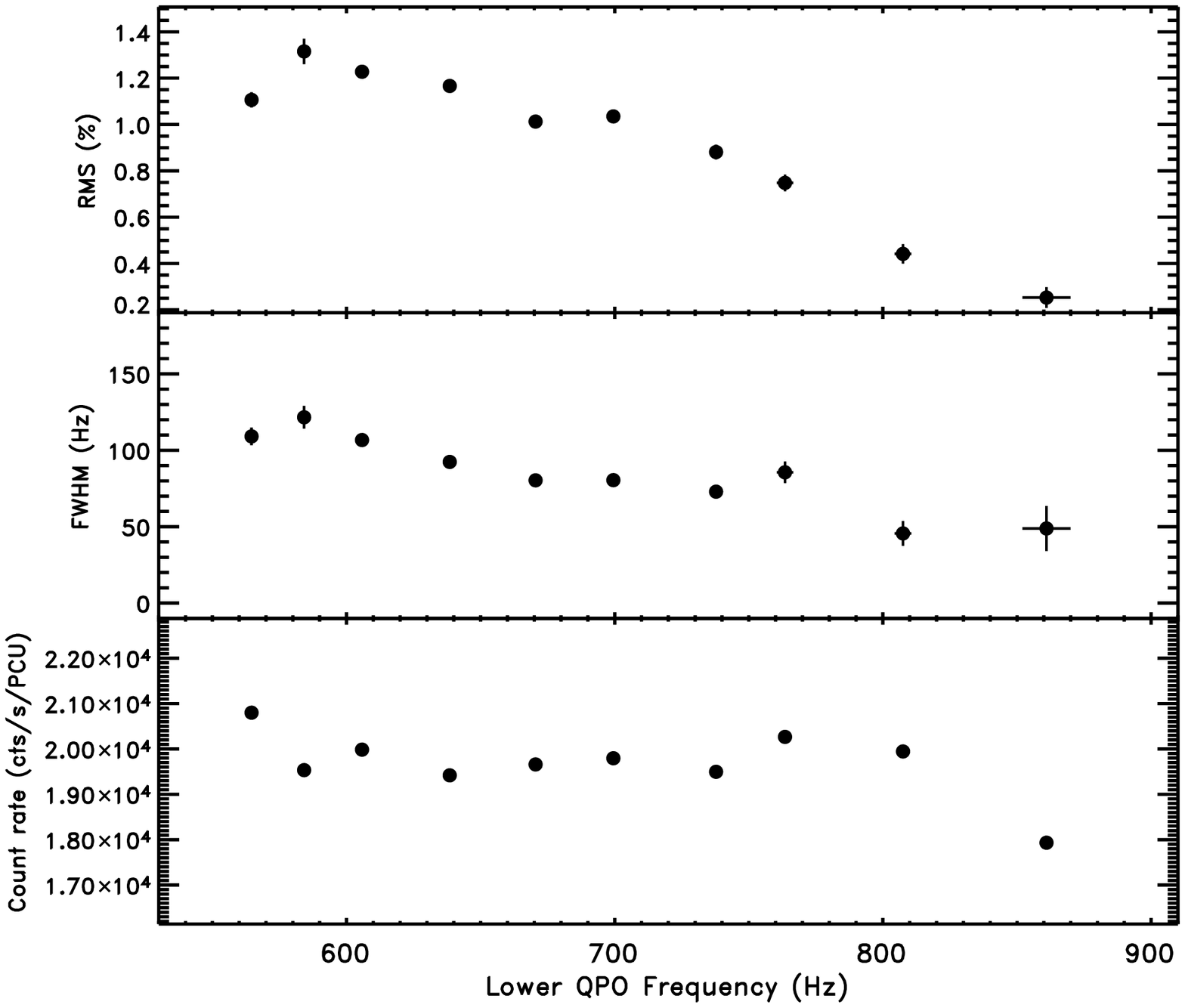}\includegraphics[width=.485\textwidth]{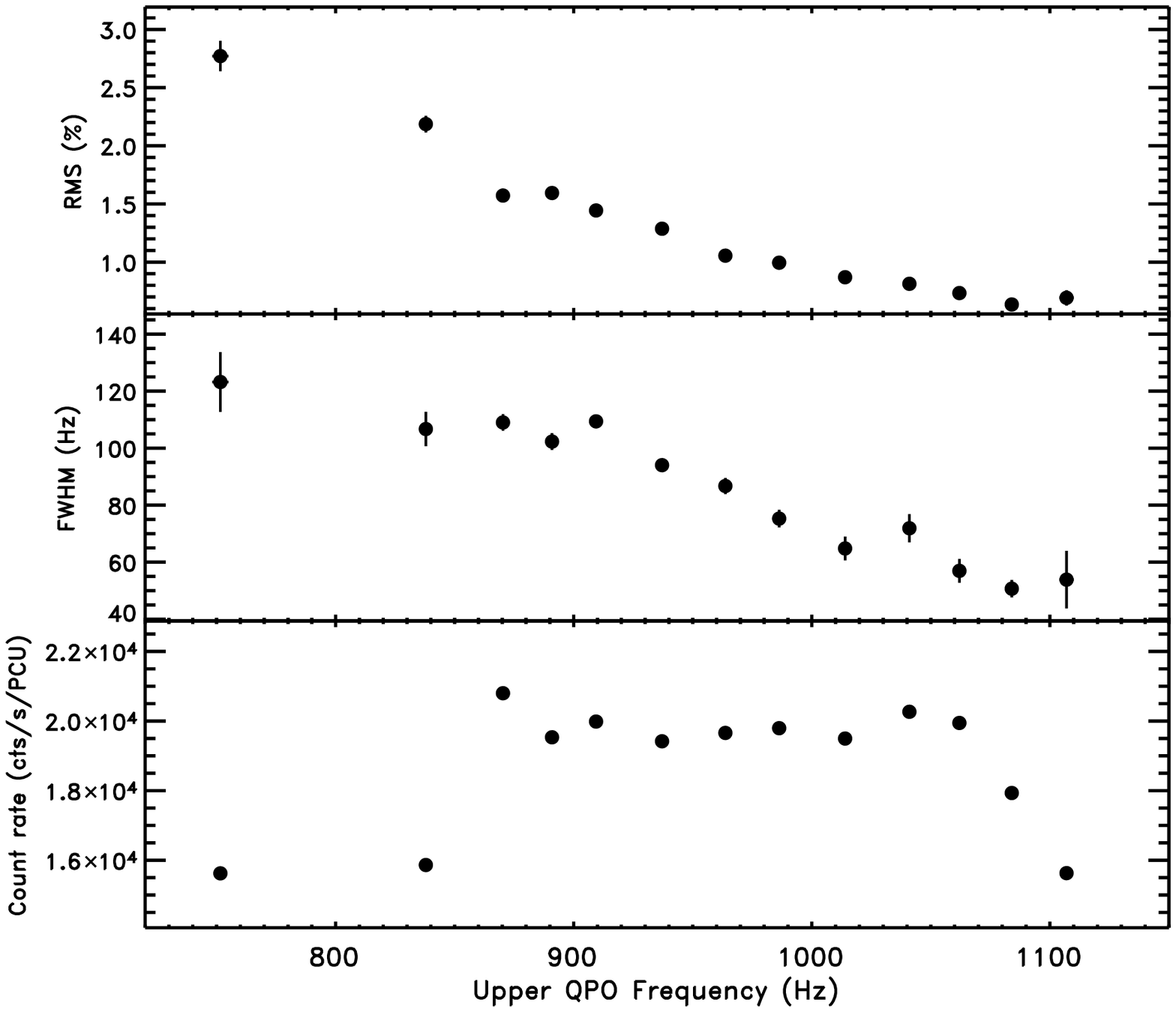}
      \caption{The RMS amplitude ($r_S$), the width ($\Delta\nu$), and the mean source count rate (normalized by the mean number of operating PCUs) for the lower  (left) and upper (right) QPO in Sco X-1, as averaged over frequency bins of 25 Hz. This is the combination of these three parameters as described in equation 1 which explains the dependence of $n_\sigma$ with frequency shown in Figure \ref{boutelier:fig7}.}
   \label{boutelier:fig8}
   \end{center}
\end{figure*}

Building on the above findings, it is worth looking back at the historical case of Sco X-1, and see whether the same effect can provide an explanation for the clustering of ratio claimed at 1.5 \citep{Abramowicz:2003sf}. For this purpose, we have reprocessed all the available data from the RXTE archive, extending on the data used by \citet{van-der-Klis:1997hb,2000MNRAS.318..938M}. We have used all science event and binned mode data with a time resolution better than 250 $\mu$s. We have again computed 16 second PDS with a 1 Hz resolution, using events recorded between 2 and 40 keV. The analysis here is complicated by deadtime effect, which we have to account for in the PDS analysis. We have modeled the modification of the PDS with a power law of index close to zero at high frequencies, and fitted the QPOs with the method described above. Over our data set, we have detected about 100 QPOs from the source on a typical ObsID timescales. Likewise for the two other sources, we identify single QPOs by placing them on a RMS/quality factor versus frequency plot. The lower QPO frequency varies from 550 Hz to 800 Hz, while the upper QPO frequency varies between 750 Hz and 1100 Hz. Previous analysis \citep{van-der-Klis:1997hb,2000MNRAS.318..938M}, based on a smaller data set, found the upper QPO frequency to range from about 850 Hz to 1150 Hz. We report for the first time, an upper QPO at a frequency of 750 Hz, detected in the ObsID 93067-01-01-02 recorded on July 3rd, 2007 at 10:33 am. Next to the 750 Hz upper QPO, the second lowest upper QPO frequency is at $ \sim 850$ Hz, implying a gap of 100 Hz in its frequency distribution. The histograms of ratios are shown in Figure \ref{boutelier:fig7}. As previously found the histograms cluster around 1.5. We have then applied the shift-and-add technique to the ObsID averaged PDS as described above, and computed the average statistical significance of both QPOs, which we then normalized to a PDS integration time of 3 kseconds. The results are also shown on Figure \ref{boutelier:fig7}. In Sco X-1, the lower QPO is detected on such a timescale only in a narrow 300 Hz interval, whereas the upper QPO is always more significant and detected over a slightly wider frequency span $\sim 350$ Hz wide (note however that there is only one ObsID in the first histogram bin where the significance reaches a maximum). Note that the narrowness of the frequency range for the both QPOs limits by itself the range of possible ratios between 1.3 and 1.6. Figure \ref{boutelier:fig8} shows how $S$, $r_S$ and $\Delta\nu$ depend on frequency for both the lower and upper kHz QPOs. As can be seen, the 750 Hz QPO follows the same trends $r_S$ and $\Delta\nu$ as upper QPOs detected above 800 Hz, and is therefore very likely a single upper QPO. Note also that the mean count rate varies by less than 20\% of the mean, across the frequency range over which both QPOs are detected.

\begin{figure}
   \begin{center}
   \includegraphics[width=.475\textwidth]{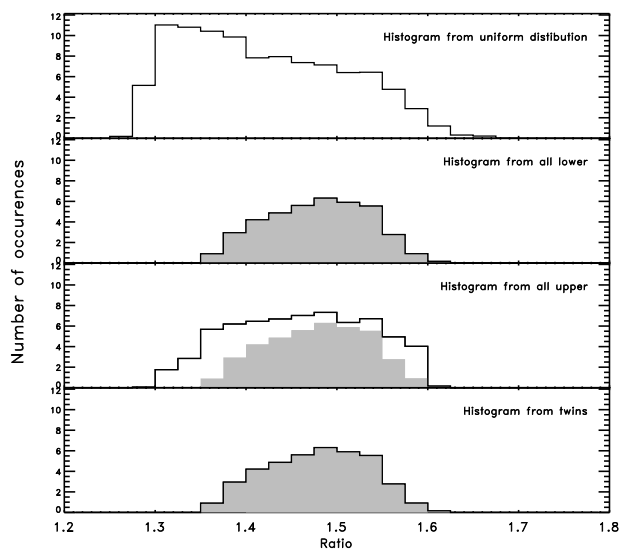}
      \caption{Simulated Sco X-1 histograms of ratios estimated from the frequencies of all single lower QPOs, all lower in twin QPOs (area in grey), all single upper QPOs, all upper in twin QPOs (area in grey) and all twin QPOs, assuming that the upper QPO frequency is uniform between 850 Hz and 1100 Hz (where most QPOs are detected). In the case of a single QPO, the frequency of the other was computed from the linear function linking the two frequencies. }
   \label{boutelier:fig9}
   \end{center}
\end{figure}

As for the two other sources, in an attempt to compare with the observations, we have simulated a uniform distribution of upper QPO frequencies between 850 Hz and 1100 Hz where the bulk of the upper QPOs are detected \citep{van-der-Klis:1997hb}. We have computed the histogram of ratios using a linear function with $\rm a=0.81$ and $\rm b=416$, which is equivalent to a power law relation $\rm \nu_{lower}=A(\nu_{upper}/1000~Hz)^B$, with A$=717$ and B$=1.78$ in the range of frequency considered. We have also included a spread in the count rate, exposure time, number of PCUs and a jitter in frequency along the correlation line. The results are shown in Figure \ref{boutelier:fig9}. Clearly, the simulations reproduces nicely the observed distributions, in particular the strong clustering of ratio around 1.5. We thus conclude that as in the two other sources, the apparent clustering of frequency ratios in Sco X-1 is also consistent with a uniform distribution of frequencies between 850 Hz and 1100 Hz. However, in Sco X-1 the clustering seems to be amplified compared to the other sources, by the fact that both QPOs can be detected simultaneously over an narrower frequency range.

\section{Conclusions}
We have shown that, because of the way the significance of kHz QPOs evolves with frequency, as a result of the QPO RMS amplitude and width being frequency dependent quantities, in sensitivity limited observations the observed clusters in frequency ratios of simultaneous twin QPO frequencies, in particular around 1.5, are consistent with an underlying  uniform frequency distribution. In other words, this implies that the clustering of ratios does not provide any evidence for preferred frequency ratios in those systems. A similar conclusion was reached by \citet{Belloni:2007ad} who found that the distribution of frequencies could be significantly different from one data set to the other. The clustering is a consequence of the fact, that albeit always present, the lower and upper QPOs can only be detected together on the same integration timescales over a limited frequency range. The clustering of ratios is more pronounced in Sco X-1 because the range of simultaneous detection of the two QPOs is narrower ($\sim 250-300$ Hz instead of 400 Hz). This conclusion applies to a sample of systems representative of all types of QPO sources. As said above, clustering of ratios caused by an incomplete sampling of the source states, whether or not its QPO frequency follows a random walk \citep{Belloni:2005vl} is more and more unlikely, as more and more data become available. 

Any models, attempting to reproduce the clusters of ratios, must therefore predict first a nearly linear relationship between the two QPO frequencies in the range where the bulk of QPOs are detected, as most models do, including non-linear resonance based models \citep{2003PASJ...55..467A}, and second, a set of well matched QPO parameters (RMS amplitude and width) which allows the two QPOs to be preferentially detected on similar timescales over limited frequency ranges. \citet{Torok:2009ay} already pointed out that at frequencies corresponding to the 3/2 ratio, the RMS amplitudes of the lower and upper QPOs as measured in the 2-40 keV band are equal (this is where the significance of the two QPOs are comparable, around 600 Hz, 900 Hz for the lower and upper respectively, see Figure \ref{boutelier:fig2}). In the framework of the non-linear resonance model proposed by \cite{2003PASJ...55..467A}, this has recently been discussed as a possible signature of a resonant energy exchange between the two QPO modes \citep{Horak:2009la}. Although such a possibility should be investigated further, it is worth stressing that non linear resonance was invoked for neutron star QPOs by interpreting the 3/2 clustering of frequency ratios as reflecting a non uniform distribution of frequencies along the line of correlation predicted by the model \citep{2003PASJ...55..467A}. This is the reason why the results presented in this paper weaken the case for a resonance mechanism at the origin of neutron star kHz QPOs, as they indicate that even a uniform distribution of frequencies will produce peaks in the ratio distribution around 3/2.

\section{Acknowledgments} It is a pleasure to thank Jean-Pierre Lasota, Mariano Mendez, M. Coleman Miller and Tomaso Belloni for detailed comments on this paper. We thank Marek Abramowicz and again Jean-Pierre Lasota for extensive discussions on the interpretation of the results. G.T. was supported by the Czech grant MSM 4781305903. We thank the referee for comments that helped to clarify the presentation of some of the main results of this paper.

\bibliographystyle{mn2e}

\end{document}